\documentclass[prl,aps,twocolumn,amsfonts,showpacs,floatfix]{revtex4}

\usepackage{epsfig}
\usepackage{psfrag}
\usepackage{color}
\usepackage{graphicx}
\usepackage{subfigure}

\begin{document}


\title{A Numerical Renormalization Group for Continuum One-Dimensional Systems}
\author{Robert M. Konik and Yury Adamov}
\affiliation{CMPMS Department, Brookhaven National Laboratory, Upton, NY 11973}

\date{\today}

\begin{abstract}
We present a renormalization group (RG) procedure which works naturally on 
a wide class of interacting one-dimension models based on perturbed (possibly strongly) continuum conformal and integrable models.
This procedure integrates Kenneth Wilson's numerical
renormalization group with Al. B. Zamolodchikov's truncated conformal spectrum
approach.
Key to the method is that such theories provide a set of completely understood eigenstates
for which matrix elements can be exactly computed.  In this procedure the RG
flow of physical observables can be studied both numerically and analytically.
To demonstrate the approach, we study the spectrum of a pair of coupled quantum Ising
chains and correlation functions in a single quantum Ising chain in the presence of a magnetic field.
\end{abstract}
\pacs{}
\maketitle

The numerical renormalization group (NRG), as developed by Kenneth Wilson \cite{wilson},
is a tremendously successful technique for the study of generic quantum impurity problems,
systems where interactions are confined to a single point.
But the NRG as such is not directly generalizable to systems where interactions are present in the bulk.
The natural generalization of the NRG in real space
treats boundary conditions between RG blocks inadequately, leading to qualitatively inaccurate
results.  To overcome this difficulty, Steven White
developed the density matrix renormalization group (DMRG) \cite{white}.  This tool is now ubiquitous
in the study of low dimensional strongly correlated {\it lattice} models and
can access both static and dynamic quantities \cite{white1}.

In this letter we offer a distinct generalization realizing a renormalization group
procedure for a wide range of strongly interacting {\it continuum} one-dimensional systems.
It can treat any model which is representable as a conformal or integrable field theory
(CFT/IFT),
with Hamiltonian, $H_0$,
plus a relevant \cite{note} perturbation (of arbitrary strength), $H_{\rm pert}$.
Beyond this there is no real constraint on $H_0$ or $H_{\rm pert}$.
In particular, the full theory, $H_0 + H_{\rm pert}$, need not be integrable or conformal.
Thus the technique can handle a standard array of models of perturbed
Luttinger liquids or Mott insulators.   It also capable of treating disordered
systems, either by envisioning $H_{\rm pert}$ as a random
field or, equally well, considering a non-unitary supersymmetric CFT, $H_0$ arising from
disorder averaging a system with quenched disorder \cite{efetov}.   This technique
also allows the study of coupled CFT/IFTs, allowing the 
study of systems between one and two dimensions.  In all cases, the low energy
spectrum and correlation functions of the model are computable.

Our starting point is the truncated spectrum approach (TSA) pioneered by Al. B. Zamolodchikov.
The TSA was developed to treat perturbations of simple conformal field theories.
While straightforward in conception, it has an advantage over other numerical techniques in that
it analytically embeds strongly correlated physics at the start, dramatically
lessening the computational burden.  In one of the TSA's first applications,
Al. B. Zamolodchikov studied a critical
Ising chain in a magnetic field \cite{zamo}, a continuum version
of the lattice model
\begin{equation}\label{ei}
H^{Ising}_0 \!=\! -J\sum_i(\sigma^z_i\sigma^z_{i+1}\!+\!\sigma^x_i); ~H_{\rm pert}\!=\!-h\sum_i\sigma^z_i,
\end{equation}
where $\sigma^{a}_i$ are the standard Pauli matrices and $i$ indexes the sites of the lattice.
The continuum model, itself integrable, has a complicated
spectrum of eight fundamental excitations.  The TSA was able to produce
the gaps of the first five excitations within 2\% of the analytic, infinite volume values by
diagonalizing a mere 39x39 matrix.  To see how remarkable this is, consider that in a 
computationally equivalent exact diagonalization of the lattice model,
one would be limited to studying a five site chain.

\begin{figure}
\includegraphics[height=1.3in,width=3.in,angle=0]{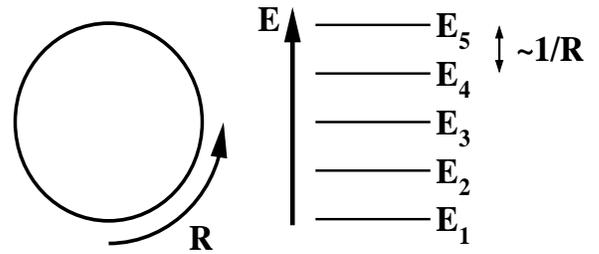}
\caption{A schematic of the finite sized system, both in real space
and in terms of energy levels, analyzed in the TSA procedure.}
\end{figure}

The TSA begins by taking the model to be studied, $H= H_0+H_{\rm pert}$, and placing
it on a finite ring of circumference, R.  Doing so makes the spectrum
discrete (see Fig. 1).   We nonetheless expect to be able to obtain {\it infinite} volume
results provided we work in a regime where $R\Delta \gg 1$ with $\Delta$ a characteristic
energy scale of the system.  In the discrete system, the spectrum can then be ordered in
energy, $|1\rangle, |2\rangle, \ldots$.  
Non-perturbative information is input in the next step of the TSA where
the matrix elements,
$H_{{\rm pert}ij} = \langle i| H_{\rm pert}|j\rangle$, are computed {\it exactly}.  It is important to
stress this is always a practical possibility.  If $H_0$ is a CFT, the attendant
Virasoro algebra (or, equally good, some more involved algebraic structure
such as a current or a W-algebra) permits the computation of such matrix elements
straightforwardly.  If $H_0$ is instead an IFT, such matrix elements
are available through the form-factor bootstrap programme \cite{ffprog}.  In an IFT, the matrix
elements which we will want to focus on involve states, $|i\rangle$, with few excitations
and, as such, are readily computable.

With the matrix elements in hand, one can then express the full Hamiltonian as
a matrix.  The penultimate step in the procedure is to truncate the spectrum
at some energy, $E_{\rm trunc}$, making the matrix finite.  This matrix is then numerically diagonalized from which
the spectrum and correlations functions can be extracted.  
When $H_0$ represents
a theory with a relatively simple set of eigenstates, this procedure, even with
a crude truncation of states, works remarkable well in extracting the spectrum.  However when the starting
point Hamiltonian, $H_0$, is more complicated (say a CFT based on a $SU(2)_k$ current algebra such as would be encountered
in the study of spin chains), 
or one is interested in computing
correlation functions in the full theory, $H_0+H_{\rm pert}$, the simple truncation scheme
ceases to produce accurate results at reasonable numerical cost. 
For example, errors in an excitation gap, $\Delta$, introduced by the truncation of states
behave as power laws, $\delta \Delta \sim E_{\rm trunc}^{-\alpha}$ ($\alpha \sim 1$).
Thus increasing $E_c$ does not dramatically reduce $\delta Q$ while
at the same time greatly increasing the cost of the exact diagonalization
routine which should scale similarly to the partition of integers, i.e. 
as $e^{\beta\sqrt{E_{\rm trunc}}}/E_{\rm trunc}$.  It is the aim of this letter
to outline an RG technique offering a dramatic improvement on this truncation scheme.

\begin{figure}[tbh]
\includegraphics[height=1.1in,width=3.3in,angle=0]{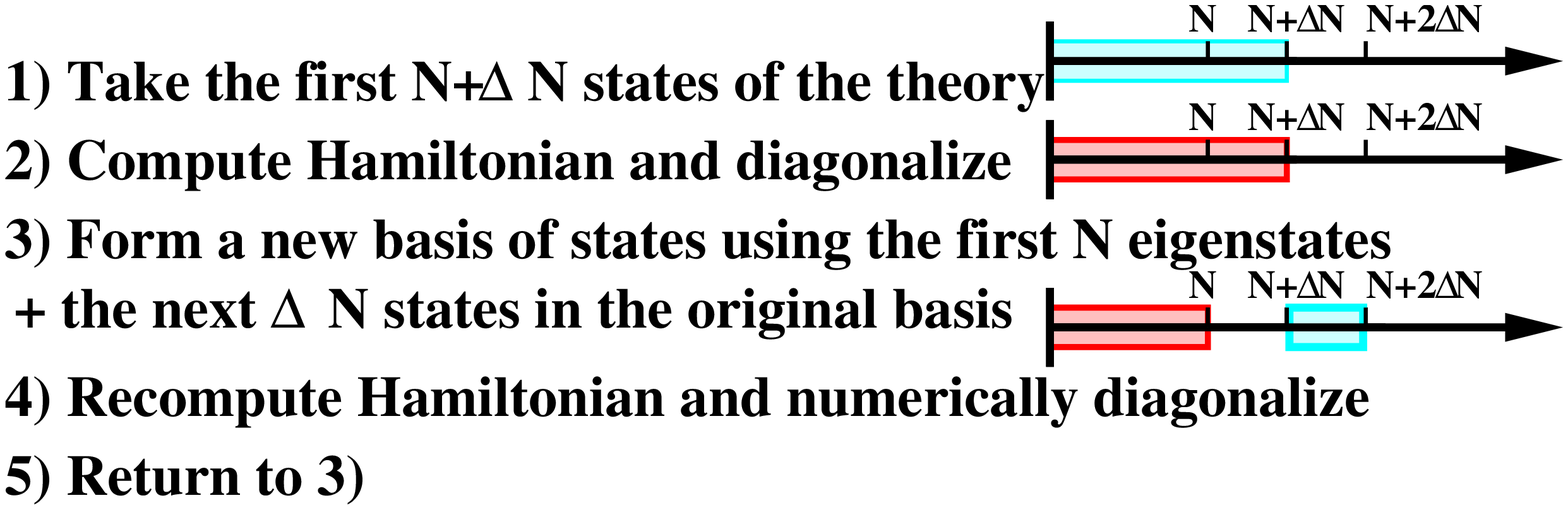}
\caption{An outline of the NRG algorithm}
\end{figure}

Our framework hews closely to the original Wilsonian conception of the NRG.  In developing
the NRG for the Kondo model, Wilson transformed the original Kondo Hamiltonian 
using the ``Kondo basis'' to a lattice model of an impurity situated at the end of a
infinite half line with sites far from the end characterized by rapidly diminishing
matrix elements.  We, in a sense, start in this position.  The ordering in energy of
the states provided by $H_0$ is in direct analogy to the half-line on which the impurity
lives in Wilson's Kondo work.  The next step in Wilson's NRG is an iterative numerical
procedure by which at each step a finite lattice is expanded by one site, the model diagonalized, and
high energy eigenstates thrown away.  
It is this iterative procedure that we mimic.  
\vskip .15in
\begin{figure}[tbh]
\includegraphics[height=2.2in,width=2.9in,angle=0]{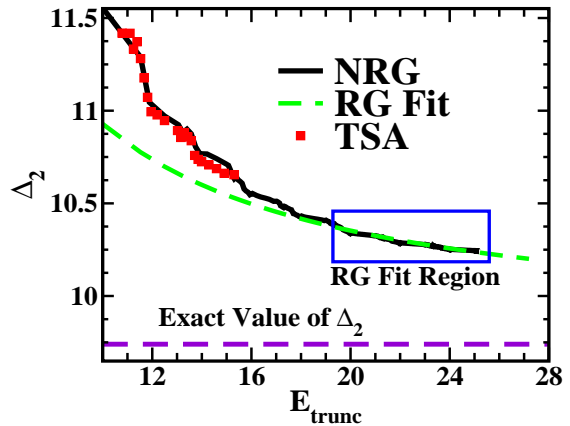}
\caption{Plots showing the behavior of $\Delta_2$ as a function
of the truncation energy, $E_{\rm trunc}$ (for $R=5\tilde\lambda^{-1}$).}
\end{figure}
\vskip -.1in
Let us denote the initial basis offered by $H_0$ as $\{|i\rangle\}^\infty_{i=1}$.
We begin by keeping a certain number, say $N+\Delta N$, of 
the lowest energy states,$\{|i\rangle\}^{N+\Delta N}_{i=1}$ (in blue in Fig. 2).   
We diagonalize the problem so extracting an initial spectrum and set
of eigenvectors, $\{|\tilde i\rangle\}^{N+\Delta N}_{i=1}$ (in red in Fig 2).  
We then toss away a certain number, $\Delta N$, of the eigenvectors corresponding to
the highest energy, i.e. $\{|\tilde i\rangle\}^{N+\Delta N}_{i=N+1}$ .  A new basis is then formed, consisting
of the remaining eigenvectors together with
the first $\Delta N$ states of $\{|i\rangle\}^\infty_{i=1}$
that we had previously ignored, 
i.e. $\{|\tilde i\rangle\}^{N}_{i=1} \cup \{|i\rangle\}^{N+2\Delta N}_{i=N+\Delta N+1}$, and the procedure
is repeated.  We present the technique schematically in Fig. 2.
Convergence of the procedure in the Kondo problem is promoted by the small matrix elements
involving sites far from the impurity.  While matrix elements in the procedure
just described grow progressively smaller under the NRG (scaling as $1/E_{trunc}$), 
here numerical convergence is not necessarily the goal.  
Rather we aim to merely bring the quantity into a regime where its flow is governed by a simple flow equation.
 
The algorithm just described implements a Wilsonian RG in reverse.   It does so at all loop orders
and so the RG flow it describes is exact.  As the flow proceeds it, however, evolves closer and closer
to a flow described by a one-loop equation.  Analytically, the equation is nearly trivial as it is given
solely in terms of the anomalous dimension, $\alpha_Q$, of the flowing quantity, $Q$ (whether
it be an energy eigenvalue or a matrix element).
More specifically,
\begin{equation}\label{eii}
\frac{d\Delta Q}{d\ln E_{\rm trunc}} = -g(\alpha_Q)\Delta Q,
\end{equation}
where $\Delta Q = Q(E_{\rm trunc})-Q(E_{\rm trunc}=\infty)$ describes the deviation of the
quantity as a function of the truncation energy from its 'true' value (i.e. the value where
the cutoff in energy is taken to infinity).  The function $g(\alpha_Q)$ can be determined exactly using
high energy perturbation theory, well-controlled
provided $H_{\rm pert}$ is relevant.  For example, if $Q$ is some energy 
eigenvalue, then $g(\alpha_Q)=\alpha_Q =1$.

The virtue of Eqn.(\ref{eii}) is that it allows us to run the RG in two steps.  We first implement the NRG
as described above until we reach a truncation energy placing us safely in the one-loop regime.
We then continue the RG by merely integrating the above equation allowing us to fully eliminate the
effects of truncation.


We now consider two examples using this RG procedure, one where we compute the spectrum of a model
and one where we analyze correlation functions.  Both examples are chosen so that a straight
application of the TSA leads to poor results.

\noindent {\bf Spectrum:} In the first example, we consider a pair of quantum critical Ising chains
coupled together:
\begin{equation}\label{eiii}
H_0 = H^{Ising~1}_0 + H^{Ising~2}_0; ~~~ H_{\rm pert}=-\lambda\int dx \sigma^z_{1}\sigma^z_{2}.
\end{equation}
This model is known to be integrable and to have a spectrum equivalent to the sine-Gordon
model at $\beta^2=\pi$ \cite{llm}, that is, a spectrum with a pair of solitons with gap, $\Delta_{s1}=\Delta_{s2}$, together with
a set of six bound states with gaps, $\Delta_k = 2\Delta_s\sin(\pi k /14)$, $k=1,\cdots,6$.
By comparing conformal perturbation theory with a thermodynamic Bethe ansatz analysis,
$\Delta_{s1,2}$ can be expressed in terms of the coupling constant, $\lambda$:
$\Delta_{s1,2} = 11.2205920 \tilde\lambda$ with $\tilde\lambda = \lambda^{4/7}/(2\pi)^{3/7}$ \cite{fateev}.

The underlying finite volume Hilbert space of $H_0$ is considerably more complicated than that of single
Ising chain.  In a single chain there are four potential sectors of the Hilbert space \cite{zamo}: 
a sector, $I$, composed of even numbers of half-integer
fermionic modes acting over a unique vacuum, $|I\rangle$; a sector, $F$, composed
of odd numbers of half-integer modes over $|I\rangle$; and finally two sectors, $\sigma/\mu$,
composed of even numbers of both right and left moving integer fermionic modes over degenerate 
vacuua, $|\sigma\rangle/|\mu\rangle$.  The sectors $\sigma$ and $\mu$ are connected
by applying a product of an odd number of even mode fermionic operators.
Under periodic boundary conditions, however, the Hilbert
space of a single chain is reduced to two sectors, $I$ and $\sigma$.  In the two chain case, this is no longer true.
Not only do we have sectors of the form $I\otimes I$, $I\otimes \sigma$, $\sigma \otimes I$,
and $\sigma\otimes\sigma$ (such tensor products arising naturally from considering
two chains), but $F\otimes F$, $F\otimes\mu$, $\mu\otimes F$, and $\mu\otimes\mu$.  Unlike
a single chain, all possible sectors
are consistent with periodic boundary conditions.
The Hilbert space that results is thus much larger and applying the TSA with a simple truncation scheme
leads to poor results for the spectrum.  Computing the spectrum of this model is thus an ideal testing ground for our proposed
RG procedure.

\begin{table}
\vskip .2in
\begin{center}
\begin{tabular}{|l|l|l|l|l|}
\hline
exc. & Exact & TSA (10) & NRG & RG Improved \\
\hline
$\Delta_{s1}$ & 11.2206 & 11.92/12.67 & 11.32/11.54  & 11.17(2)/11.15(5)\\
\hline
$\Delta_{s2}$ & 11.2206 & 11.92/12.66 & 11.32/11.54 & 11.17(2)/11.16(6)\\
\hline
$\Delta_1$ & 4.9936 & 5.29/5.61 & 5.03/5.12 & 4.97(1)/4.97(2)\\
\hline
$\Delta_2$ & 9.7369 & 10.69/11.55 & 9.89/10.24 & 9.70(3)/9.7(1)\\
\hline
$\Delta_3$ & 13.9918 & 15.58/16.65 & 14.33/14.84 & 14.02(5)/14.20(5)\\
\hline
$\Delta_4$ & 17.5452 & 19.672/20.923 & 18.69/18.03 & 17.6(1)/17.7(1)\\
\hline
$\Delta_5$ & 20.2188 & 23.64/24.64 & 20.80/21.62 & 20.2(2)/20.5(2)\\
\hline
$\Delta_6$ & 21.8785 & 23.65/25.28 & 22.39/23.08 & 21.8(1)/21.8(2)\\
\hline
\end{tabular}
\caption{The excitation energies for two coupled Ising chains at values of $R=4/5\tilde\lambda^{-1}$.
(in units of $\tilde\lambda=\lambda^{4/7}/(2\pi)^{3/7}$).}
\end{center}
\vskip -.1in
\end{table}

In Figure 3 we outline the procedure by which we extract the values of the spectrum of the two
coupled Ising chains focusing for specificity on the second bound state, $\Delta_2$,
in the spectrum.  We first show the results of a straight TSA analysis (red squares) as a
function of increasing truncation energies (given in units of $1/R$).  While the gap, $\Delta_2$, is
converging towards its infinite volume value, $9.7368648\ldots\tilde\lambda$, it is doing so only slowly and
at exponentially increasing numerical cost.  We have performed the straight TSA analysis
up to level 15 (i.e. keeping states with energies less than $15/R$) where the Hilbert space contains $\approx 3500$
states.  At this point, the TSA produces a result deviating by $\approx 20\%$ from the exact value.
We also plot the value of $\Delta_2$ as 
given by the NRG algorithm as it iterates through states of ever higher energy.
Here we have run the algorithm so as to take into account states up to level 25 (in total $\approx 150000$/sector).
We see, reassuringly, that where the TSA results exists, the NRG algorithm produces matching
results (at a fraction of the numerical cost).  The NRG algorithm ends up producing a value of $\Delta_2$
with a $5\%$ error.
Finally we plot 
the result of fitting Eqn.(\ref{eii}) to the NRG results between level 20 and 25 (where we believe
a one-loop RG equation describes the NRG flow).  Extrapolating the fit to $E_{\rm trunc}=\infty$
gives  $\Delta_2=9.70\tilde\lambda$, a value deviating from the exact result by $0.5\%$.

In Table 1 we present the results of our RG analysis on the complete spectrum of the two coupled chains.  
In the first column we provide the exact value of spectrum as determined by integrability
and the TBA analysis of Ref. \cite{fateev}.  In the second column we give the values of the spectrum
at two different system sizes ($R=4/5\tilde\lambda^{-1}$)
computed using a straight TSA analysis truncating at level 10 (i.e. keeping $\approx 600$ states
in each of the relevant sectors).  
We see that the disagreement  with the exact result ranges up to 20\%.
In the third column we give the results coming from applying the NRG algorithm (again iterating
until we have reached level 25).
We see a marked improvement over the TSA analysis, but still we obtain results with errors 
ranging up to 5\%. In the final column, we give the results for the spectrum arrived
at by fitting the one-loop RG equation (Eqn.(\ref{eii})) to the NRG data.  We see that our errors are
now less than 1\% .
\vskip .1in
\begin{figure}[tbh]
\includegraphics[height=2.4in,width=3.in,angle=0]{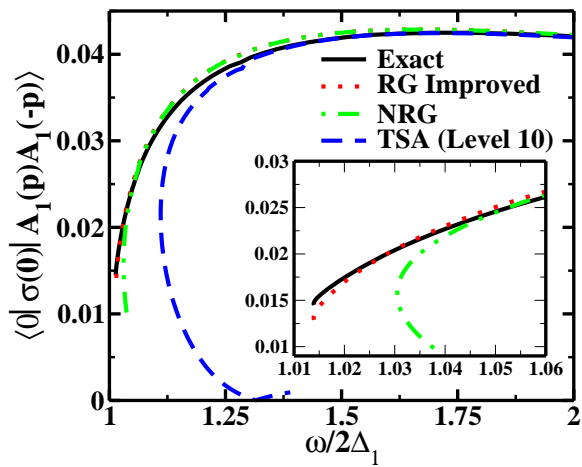}
\caption{A plot of the matrix element, $\langle 0| \sigma(0) |A_1(p)A_1(-p)\rangle$
as a function of energy, $\omega = 2(p^2+m_1^2)^{1/2}$.}
\end{figure}

\noindent {\bf Correlation Functions:}
We now turn to the computation of correlation functions using the above described RG methodology.  For simplicity
we consider only $T=0$ response functions although a generalization to finite temperature multi-point 
functions is readily realizable.
At zero temperature, the imaginary piece of a retarded correlation function, $G_{\rm ret}(x,t) = 
\langle {\cal O}(x,t){\cal O}(0)\rangle_{\rm ret}$,
has a spectral decomposition, $S_{\cal OO}(x,\omega) = -\frac{1}{\pi}{\rm Im}G_{\rm ret}(x,\omega>0)$ equal to \cite{review}
\begin{eqnarray}\label{eiv}
S_{\cal OO}(x,\omega) = \sum_{n,s_n}e^{iP_{s_n}x}|\langle 0|{\cal O}(0,0)|n;s_n\rangle|^2
\delta(\omega-E_{s_n}),
\end{eqnarray}
where $|n;s_n\rangle$ is an eigenstate of the system with energy/momentum $E_{s_n}/P_{s_n}$ 
built out of $n$ fundamental single
particle excitations carrying internal quantum numbers $s_n$.
Thus the computation of any response function is equivalent to the computation of a number
of matrix elements of the form $\langle 0|{\cal O}(x,0)|n;s_n\rangle$.  Ostensibly 
to compute the response function fully one would need to compute an infinite number of such
matrix elements.  In practice if one is interested in the response function at low energies
only a small finite number of such matrix elements need be computed \cite{review}.
We will illustrate the computation under the RG of one such non-trivial matrix element
for a critical Ising chain in a magnetic field (i.e. Eqn.(\ref{ei})). 
While the spectrum of this theory rapidly converges upon the increase of $E_{\rm trunc}$,
the matrix elements are less well behaved.  And unlike the two-chain case, analytical
results are available for comparison \cite{muss}. Thus this computation is a good test
of our RG methodology.

We specifically compute the two excitation contribution 
$f_2(\omega) = \langle0|\sigma(0)|A_1(p)A_1(-p)|0\rangle$ with
$p = (\omega^2-\Delta_{1I}^2)$, to
the spin-spin correlator, $S_{\sigma\sigma}(p=0,\omega>0)$.  (Here $\Delta_{1I}$ is
the gap of the lowest lying single particle excitation, $A_1$, in an Ising chain in a magnetic field.)
This contribution takes
the form
\begin{equation}
\delta S_{\sigma\sigma}(\omega) = \pi^{-1}\omega(\omega^2-4\Delta_{1I}^2)^{1/2}\Theta(\omega-2\Delta_{1I})
|f_2(\omega)|^2.
\end{equation}
We are able to compute the necessary {\it infinite} volume matrix element over a {\it continuous} range of energies
by studying a single matrix element in {\it finite} volume where the spectrum is {\it discrete}.  We do so by
continuously varying the system size, $R$.  Under such variations, the energy $\omega = 2(p^2+m_1^2)^{1/2}$ 
of the state, $|A_1(p)A_1(-p)\rangle$, changes continuously due to the quantization condition of the momentum, $p$,
(i.e. $p = 2\pi n /R + \delta(p,-p)$ where $\delta(p,-p)$ is a two-body scattering phase).  

In Figure 4 by varying $R$ we parametrically plot the results of our computations of $f_2(p)$ vs its exact value \cite{muss}.
A straight application of the TSA (with a level
10 truncation) produces acceptable results at higher energies but does poorly at energies around threshold, $2\Delta_1$.
At larger values of $R$ (and so smaller energies), the TSA breaks down.  The TSA curve
in this region is then double valued.  Computing the same matrix
element with the NRG algorithm leads to a considerable improvement but at the lowest energies
a deviation from the exact result remains (see inset to Fig. 4).  
RG improving the computation of $f_2(p)$ largely removes this discrepancy even at energies next
to threshold.  (In applying Eqn. (2) to $Q=f_2(p)$, perturbation theory yields, $g(\alpha_{f_2})= 2(1-1/8)$ where $1/8$
is the anomalous dimension of the spin operator, $\sigma$).

In conclusion we have presented an RG scheme by which a large number of one-dimensional
continuum models can be studied with quantitative accuracy.  With this methodology, both
the spectrum and spectral functions of a model can be determined.

RMK and YA acknowledge support from the US DOE
(DE-AC02-98 CH 10886) together with useful discussions with A. Tsvelik.


\begin{references}
\bibitem{note} Further studies need to be done on the efficacy of the method for marginal perturbations.
\end{references}
\end{document}